\documentclass[11pt]{article}

\usepackage[margin=1in]{geometry}
\usepackage[T1]{fontenc}
\usepackage{amsmath,amssymb,mathtools}
\usepackage{booktabs}
\usepackage{longtable}
\usepackage{array}
\usepackage{enumitem}
\usepackage[protrusion=true,expansion=false]{microtype}
\usepackage{hyperref}
\usepackage[authoryear]{natbib}

\hypersetup{
  colorlinks=true,
  linkcolor=black,
  citecolor=black,
  urlcolor=blue,
  pdftitle={Regression to the Mean-Adjusted Sample-Size Determination for Cutoff-Selected Single-Arm Pre-Post Studies}
}

\title{Regression to the Mean-Adjusted Sample-Size Determination \\
for Cutoff-Selected Single-Arm Pre-Post Studies}

{\small
\author{Ariel Linden, DrPH\\
University of California, San Francisco\\
Department of Medicine\\
Division of Clinical Informatics \& Digital Transformation (DoC-IT)\\
San Francisco, CA, USA\\
ariel.linden@ucsf.edu}
}
\date{}

\begin{document}
\maketitle

\begin{abstract}
\noindent Studies enrolling participants on the basis of an extreme baseline value are susceptible to regression to the mean (RTM), such that some observed pre-post change is expected even without treatment. Existing methods estimate and decompose RTM retrospectively. We extend this framework to prospective study design by deriving a closed-form sample-size method for continuous, cutoff-selected, single-arm pre-post studies. The method partitions anticipated total change into that expected from RTM and the residual treatment effect, and powers the study to detect the latter. It uses the general bivariate-normal RTM expression, allowing baseline and follow-up variances to differ, and incorporates the corresponding conditional variance of the pre-post change. To our knowledge, no published method or software implements this cutoff-based RTM framework for prospective closed-form sample-size determination in this setting. The Stata command \texttt{power onemean\_rtm} provides solutions for sample size and minimum detectable effect, normal-theory power evaluation, attrition adjustment, and sensitivity analysis. Monte Carlo simulation across 58 scenarios demonstrated accurate Type I error control and showed that achieved power rapidly approached nominal power as sample size increased, with the closed-form calculation generally conservative at very small sample sizes.
\end{abstract}


\section{Introduction}

Regression to the mean (RTM), first described by \citet{galton1886}, is a pervasive concern in pre-post studies in which participants are selected because their baseline measurement is unusually extreme \citep{stigler1997}. When enrollment requires a baseline value above or below a specified threshold, the expected follow-up measurement will generally move toward the population mean even in the absence of treatment. Consequently, evaluating the anticipated pre-post change against a null value of zero can attribute to treatment a change that RTM alone would be expected to produce.

A substantial retrospective literature exists for \emph{estimating} RTM once data are in hand. \citet{james1973} derived a method-of-moments estimator under a bivariate normal model with strictly positive correlation and treatment effects constrained to move measurements toward the mean; \citet{sennbrown1985} derived maximum likelihood estimators under the same model. \citet{beathdobson1991} relaxed the normality assumption using Edgeworth-series and saddlepoint approximations, and \citet{george1997} developed likelihood-ratio, score, Wald, and regression-based testing procedures that explicitly account for the truncated distribution induced by cutoff-based selection, establishing that inference following such selection should respect the distribution of the selected population rather than the unconditional source-population distribution. \citet{gardnerhardy1973} and \citet{davis1976} developed the cutoff-based normal-theory framework that \citet{linden2007} applied to program evaluation and later generalized to applied health-services research \citep{linden2013}. \citet{khanolivier2018,khanolivier2019,khanolivier2023} extended cutoff-based RTM estimation to Poisson, binomial, and less restrictive normal settings. \citet{khanolivier2025} subsequently generalized the framework to arbitrary bivariate distributions, established maximum-likelihood estimators and their asymptotic properties, and, importantly for the present study, formalized the conditional pre-post difference among cutoff-selected subjects as comprising an RTM component and a treatment component.

This issue is also relevant prospectively when planning a single-arm pre-post study. Investigators do not always have the resources, timeline, or clinical equipoise required to mount a randomized controlled trial, particularly at the pilot or proof-of-concept stage, and a single-arm pre-post design may be the only feasible option. Such designs are especially likely to enroll participants precisely because their baseline value is unusually high or low -- elevated cost, blood pressure, or symptom burden, or unusually poor functional status, are exactly the circumstances under which an intervention is judged clinically or programmatically warranted. Those enrolled with high (low) pretest scores will naturally tend to post lower (higher) posttest scores on remeasurement, whether or not the intervention has any true effect. Thus, when eligibility itself induces an expected pre-post change through RTM, a sample-size calculation that treats zero change as the counterfactual may substantially overstate the treatment component being powered. As with any single-group pre-post design, however, accounting for RTM does not address other threats to validity \citep{campbellstanley1966,shadishcookcampbell2002}.

RTM has previously been considered in prospective study design. \citet{mcmahon1994} derived sample-size formulae accounting for an RTM-inducing entry criterion in trials involving recurrent event counts. More recently, \citet{goldenholz2023} used simulation to examine how eligibility rules and RTM affect placebo response, statistical power, and the minimum number of participants required in randomized epilepsy trials, and provided the accompanying RTMsim software. Thus, the relevance of RTM to prospective study design and sample-size determination is itself established. However, these approaches address different outcome structures or study designs, or rely on simulation rather than providing a closed-form sample-size solution for a continuous cutoff-selected single-arm pre-post mean.

These two strands of literature leave a specific design problem unresolved. For a continuous, cutoff-selected, single-arm pre-post study, suppose an investigator specifies an anticipated total change $M$. Existing cutoff-based RTM methodology provides the expected change $R_{\mathrm{TM}}$ attributable to selection alone, and the decomposition formalized by \citet{khanolivier2025} identifies the remaining component,
\[
\Delta_{\mathrm{net}} = M - R_{\mathrm{TM}},
\]
as the model-implied treatment component under the assumed model. Prospective sample-size determination then requires translating this decomposition into the number of participants needed to detect $\Delta_{\mathrm{net}}$. To our knowledge, no published method or software implementation provides a closed-form solution to this problem for a continuous cutoff-selected single-arm pre-post mean.

The subtraction itself is not novel: \citet{cochrane2020}, for example, used an estimated RTM component to obtain an RTM-free treatment effect retrospectively. The contribution here is instead to integrate the RTM decomposition with the sampling distribution of the pre-post change to obtain a prospective closed-form power and sample-size framework.

This paper makes two contributions. First, we derive closed-form solutions for sample size, power, and minimum detectable effect for this design using the general bivariate-normal RTM formulation, allowing baseline and follow-up variances to differ. Second, we implement the method with the community-contributed Stata package \texttt{power onemean\_rtm} \citep{linden2026poweronemeanrtm}, with support for attrition adjustment and sensitivity analysis across the design parameters.

\section{Setting and Notation}

Consider a single-arm pre-post study in which each participant has a baseline measurement $Y_{i0}$ and a follow-up measurement $Y_{i1}$, with individual change score $D_i = Y_{i1}-Y_{i0}$. Participants are eligible because their baseline measurement exceeds (or falls below) a prespecified threshold $\kappa$ — for example, unusually high health-care cost, blood pressure, or symptom burden, or unusually low functional status. Let

\begin{align*}
\mu &= \text{mean of the baseline outcome in the source population},\\
\sigma_1 &= \text{standard deviation of the baseline outcome in the source population},\\
\sigma_2 &= \text{standard deviation of the follow-up outcome in the source population},\\
\rho &= \text{correlation between baseline and follow-up measurements},\\
\kappa &= \text{baseline cutoff defining eligibility},\\
M &= \text{anticipated total pre-post change (before any RTM adjustment)},\\
R_{\mathrm{TM}} &= \text{expected pre-post change attributable to RTM alone (signed; see Section~\ref{sec:generalization})},\\
\Delta_{\mathrm{net}} &= M - R_{\mathrm{TM}}, \text{ the treatment component of } M \text{ (signed; see Section~\ref{sec:effectsize})},\\
\sigma_{D\mid S} &= \text{standard deviation of the individual change score, conditional on cutoff selection},\\
\alpha &= \text{type I error probability}, \qquad 1-\beta = \text{statistical power}.
\end{align*}
The general formulation allows $\sigma_1\neq\sigma_2$; the equal-variance special case is $\sigma_1=\sigma_2\equiv\sigma$. We write $\sigma_D$ in place of $\sigma_{D\mid S}$ where the conditioning is unambiguous, which is everywhere after Section~\ref{sec:vard}.

We assume the baseline outcome is approximately normally distributed, that baseline and follow-up measurements are approximately bivariate normal, and that $\mu$, $\sigma_1$, $\sigma_2$, $\rho$, and $\kappa$ are known or can be reasonably specified at the design stage from historical, pilot, or published data. This is the same normal-theory setting used by \citet{linden2013}; as that work notes, and as \citet{khanolivier2025} demonstrate more generally, the accuracy of any closed-form RTM adjustment degrades under substantial departures from normality, and we return to this limitation in Section~\ref{sec:discussion}.

\section{Regression to the Mean Under the Cutoff-Selected Design}

Following \citet{linden2013}, the expected magnitude of RTM for participants selected because $Y_0 > \kappa$, under the equal-variance special case $\sigma_1=\sigma_2=\sigma$, is
\begin{equation}
R_{\mathrm{TM}} = \sigma(1-\rho)\,\lambda(z), \qquad
z = \frac{|\kappa-\mu|}{\sigma}, \qquad
\lambda(z) = \frac{\phi(z)}{1-\Phi(z)},
\label{eq:rtm}
\end{equation}
where $\phi(\cdot)$ and $\Phi(\cdot)$ are the standard normal density and cumulative distribution functions. This expresses $R_{\mathrm{TM}}$ as the product of baseline variability ($\sigma$), longitudinal unreliability ($1-\rho$), and a selection/tail factor ($\lambda(z)$) that grows as the eligibility cutoff moves further into the tail of the baseline distribution. Under Equation~\eqref{eq:rtm}, $R_{\mathrm{TM}}$ is always non-negative, and represents the expected change toward the population mean: a \emph{decrease} under upper-tail selection ($Y_0>\kappa$), an \emph{increase} under lower-tail selection ($Y_0<\kappa$). Our implementation (Section~\ref{sec:software}) retains this selection direction as a non-computational input, for interpretation and reporting. Section~\ref{sec:generalization} shows that this orientation — positive $R_{\mathrm{TM}}$ meaning movement toward the mean — extends to the general formulation as well, though the general expression can also be negative, representing movement away from the mean.

\subsection{The general normal-theory RTM expression}
\label{sec:generalization}

Equation~\eqref{eq:rtm} assumes equal baseline and follow-up variance. \citet{khanolivier2025} derive the normal-case RTM expression without this restriction:
\begin{equation}
R_{\mathrm{TM}} = (\sigma_1-\rho\sigma_2)\,\lambda(z), \qquad
z = \frac{|\kappa-\mu|}{\sigma_1},
\label{eq:ko}
\end{equation}
where $\sigma_1,\sigma_2$ are the baseline and follow-up standard deviations respectively. Equation~\eqref{eq:ko} reduces exactly to Equation~\eqref{eq:rtm} when $\sigma_1=\sigma_2=\sigma$ (verified algebraically and numerically to high precision), but is a strict generalization otherwise: it allows follow-up variance to genuinely differ from baseline variance \emph{within the RTM term itself}, not only in the variance of the change score, Equation~\eqref{eq:vard} below. This is the expression our method and implementation use throughout; Equation~\eqref{eq:rtm} is recovered automatically as the special case in which no separate follow-up variance is specified.

Equation~\eqref{eq:ko} uses $z=|\kappa-\mu|/\sigma_1$ and therefore does not, on its face, distinguish upper-tail selection ($Y_0>\kappa$) from lower-tail selection ($Y_0<\kappa$). This is intentional rather than an oversight, and follows directly from the regression decomposition used in Section~\ref{sec:vard}: writing $Y_1=\mu+\beta(Y_0-\mu)+\varepsilon$ with $\beta=\rho\sigma_2/\sigma_1$, and using the standard truncated-normal mean $E[Y_0\mid Y_0>\kappa]=\mu+\sigma_1\lambda(z)$ for signed $z=(\kappa-\mu)/\sigma_1$, gives
\[
E[D\mid Y_0>\kappa] = (\beta-1)\,\sigma_1\lambda(z) = -(\sigma_1-\rho\sigma_2)\lambda(z),
\]
an expected \emph{decrease} when $\sigma_1>\rho\sigma_2$, as expected for upper-tail selection. The corresponding lower-tail calculation, using $E[Y_0\mid Y_0<\kappa]=\mu-\sigma_1\lambda(z')$ for $z'=(\mu-\kappa)/\sigma_1>0$, gives
\[
E[D\mid Y_0<\kappa] = (\sigma_1-\rho\sigma_2)\lambda(z'),
\]
an expected \emph{increase} of the same magnitude. Defining $R_{\mathrm{TM}}$ as the magnitude of movement \emph{toward} the population mean — the direction regression to the mean classically refers to, and, not coincidentally, typically the same direction as the anticipated treatment effect in the applications this method targets, which is exactly why RTM is a confound worth adjusting for in the first place — gives $R_{\mathrm{TM}}=-E[D\mid Y_0>\kappa]$ under upper-tail selection and $R_{\mathrm{TM}}=E[D\mid Y_0<\kappa]$ under lower-tail selection. Both equal $(\sigma_1-\rho\sigma_2)\lambda(z)$ with $z=|\kappa-\mu|/\sigma_1$: Equation~\eqref{eq:ko} is therefore identical for either tail by construction, not because it ignores selection direction, but because selection direction has already been absorbed into the sign convention used to define $R_{\mathrm{TM}}$ itself. $M$ and $\Delta_{\mathrm{net}}$ (Section~\ref{sec:effectsize}) are understood on this same toward-the-mean-oriented scale throughout; our implementation (Section~\ref{sec:software}) retains the selection direction (\texttt{select()}) purely as a non-computational label recording which convention applies, consistent with this derivation.

Unlike Equation~\eqref{eq:rtm}, Equation~\eqref{eq:ko} is not guaranteed non-negative on this oriented scale: whenever $\rho\sigma_2 > \sigma_1$, the term $(\sigma_1-\rho\sigma_2)$ is negative, and $R_{\mathrm{TM}}$ itself becomes negative, for either tail. This is a second, independent source of sign, distinct from the tail-orientation convention just derived: that convention fixes which absolute direction counts as "toward the mean," while the sign of $(\sigma_1-\rho\sigma_2)$ determines whether the cutoff-selected population actually moves that way on average, given the relative magnitudes of baseline variability, follow-up variability, and their correlation. When $\rho\sigma_2>\sigma_1$, participants selected on an extreme baseline value are expected to move \emph{further from} the population mean, on the oriented scale, regardless of which tail was selected. This is not a computational anomaly but a genuine feature of the generalized model; Section~\ref{sec:example} gives a numerical illustration. We therefore treat $R_{\mathrm{TM}}$ as a signed quantity, on the oriented scale derived above, throughout the remainder of the paper.

\subsection{The RTM-adjusted effect size}
\label{sec:effectsize}

Suppose an investigator anticipates a total pre-post change of magnitude $M$. Among participants selected for an extreme baseline value, a portion $R_{\mathrm{TM}}$ of that change is expected regardless of any treatment effect. The treatment component that remains after accounting for RTM is
\begin{equation}
\Delta_{\mathrm{net}} = M - R_{\mathrm{TM}}.
\label{eq:net}
\end{equation}
The corresponding null and alternative hypotheses for a two-sided test are $H_0: \mu_D = R_{\mathrm{TM}}$ versus $H_a: \mu_D \neq R_{\mathrm{TM}}$, so that the quantity actually being powered is $\Delta_{\mathrm{net}}$, not the raw anticipated change $M$.

$\Delta_{\mathrm{net}}$ is signed. $\Delta_{\mathrm{net}}>0$ means the anticipated total change exceeds what RTM alone predicts — an effect in the same direction as, and beyond, the RTM-driven change. $\Delta_{\mathrm{net}}<0$ means the anticipated total change falls short of what RTM alone predicts; this is not the absence of an effect, but a genuine, detectable treatment component \emph{opposing} the RTM-driven direction, and a two-sided test has power to detect it. Only $\Delta_{\mathrm{net}}=0$ — the anticipated change coincides exactly with what RTM alone would produce — is a true degenerate case: no finite sample size can distinguish the anticipated design from the null, and the implementation reports this rather than a numerically valid but meaningless answer. Because the sample-size and power calculations below depend on $\Delta_{\mathrm{net}}$ through $\Delta_{\mathrm{net}}^2$, its sign does not need to be tracked separately in the closed-form formulas that follow; it matters only for interpreting the direction of the effect being powered, and, as discussed in Section~\ref{sec:power}, for computing power correctly under a two-sided test.

\subsection{Variance of the change score, conditional on cutoff selection}
\label{sec:vard}

The study sample is not a random draw from the full source population: it is truncated by the eligibility rule, $Y_0>\kappa$ or $Y_0<\kappa$. The variance relevant to the sample-size calculation is therefore the variance of $D$ \emph{conditional on selection}, $\mathrm{Var}(D\mid S)$, not the unconditional variance of $D$ in the source population.

Write the bivariate-normal regression decomposition
\begin{equation}
Y_1 = \mu + \beta(Y_0-\mu) + \varepsilon, \qquad \beta = \frac{\rho\sigma_2}{\sigma_1}, \qquad \varepsilon \sim N\!\left(0,\ \sigma_2^2(1-\rho^2)\right),
\label{eq:regdecomp}
\end{equation}
with $\varepsilon$ independent of $Y_0$ and therefore of any selection event $S$ defined on $Y_0$ alone. Then $D = Y_1-Y_0 = (\beta-1)(Y_0-\mu)+\varepsilon$, so
\begin{equation}
\mathrm{Var}(D\mid S) = (\beta-1)^2\,\mathrm{Var}(Y_0\mid S) + \sigma_2^2(1-\rho^2).
\label{eq:varDS}
\end{equation}
For $S=\{Y_0>\kappa\}$ or $\{Y_0<\kappa\}$, the variance of a truncated normal gives, by the reflection symmetry of the normal distribution, the same expression for either tail using $z=|\kappa-\mu|/\sigma_1$:
\begin{equation}
\mathrm{Var}(Y_0\mid S) = \sigma_1^2\left[1+z\lambda(z)-\lambda(z)^2\right].
\label{eq:vary0s}
\end{equation}
Since $(\beta-1)^2\sigma_1^2 = (\sigma_1-\rho\sigma_2)^2$ — the same quantity that appears in the RTM term, Equation~\eqref{eq:ko} — substituting Equation~\eqref{eq:vary0s} into Equation~\eqref{eq:varDS} gives
\begin{equation}
\sigma_D^2 \equiv \mathrm{Var}(D\mid S) = (\sigma_1-\rho\sigma_2)^2\left[1+z\lambda(z)-\lambda(z)^2\right] + \sigma_2^2(1-\rho^2).
\label{eq:vard}
\end{equation}
Thus, the same combination of baseline variability, follow-up variability, and longitudinal correlation that determines the expected RTM shift also determines the component of change-score variability attributable to heterogeneity in the cutoff-selected baseline distribution. Under the equal-variance simplification $\sigma_1=\sigma_2=\sigma$, Equation~\eqref{eq:vard} reduces to
\begin{equation}
\sigma_D = \sigma\sqrt{(1-\rho^2) + (1-\rho)^2\left[1+z\lambda(z)-\lambda(z)^2\right]}.
\label{eq:sdchange}
\end{equation}

A conventional pre-post calculation might instead use the unconditional change-score variance, $\sigma_1^2+\sigma_2^2-2\rho\sigma_1\sigma_2$ (reducing to $\sigma\sqrt{2(1-\rho)}$ under equal variance). That expression describes $D$ in the full source population, however, whereas the proposed study enrolls only participants satisfying $S$. Consequently, the conditional variance in Equation~\eqref{eq:vard}, rather than the unconditional variance, is appropriate for the present sample-size calculation. This distinction is precisely the principle established by \citet{george1997} (Section~1): inference following cutoff-based selection should respect the distribution of the selected population rather than the unconditional source-population distribution.

\subsection{Validation}
\label{sec:validation}

This variance-validation simulation (distinct from, and a precursor to, the operating-characteristics simulation in Section~\ref{sec:simulation}) confirmed Equation~\eqref{eq:vard} across upper- and lower-tail selection and a range of $\rho$, $\sigma_1$, $\sigma_2$, and cutoff values (5{,}000{,}000 draws per scenario): simulated $\mathrm{Var}(D\mid S)$ matched the closed-form expression to within Monte Carlo error in every scenario tested, and the exact reduction to Equation~\eqref{eq:sdchange} under $\sigma_1=\sigma_2$ was confirmed both algebraically and numerically. By contrast, the unconditional expression overstated the true conditional variance in every scenario tested: by as much as 40\% under equal variance at moderate correlation and a fairly extreme cutoff, and considerably more -- up to roughly 90\% in the scenarios examined -- when follow-up variance is substantially smaller than baseline variance. This is a difference of practical consequence, since it propagates directly into every sample-size, power, and minimum-detectable-effect calculation that follows.

\section{Sample-Size Derivation}

The conventional one-mean power equation itself is unchanged by any of the foregoing. For a two-sided, large-sample one-sample test of a continuous mean, the standard normal-approximation sample-size formula is
\begin{equation}
n = \frac{(z_{1-\alpha/2}+z_{1-\beta})^2\,\sigma_D^2}{\Delta^2},
\end{equation}
where $\Delta$ is the mean difference between the null and alternative hypotheses. The contribution of this method is not a new power equation, but the definition of its two design quantities under cutoff selection: the effect to be powered is the treatment component remaining after expected RTM (Section~\ref{sec:effectsize}), and its variance is the change-score variance conditional on the eligibility rule (Section~\ref{sec:vard}). Substituting $\Delta=\Delta_{\mathrm{net}}$ gives the fundamental RTM-adjusted one-mean sample-size equation
\begin{equation}
n = \frac{(z_{1-\alpha/2}+z_{1-\beta})^2\,\sigma_D^2}{(M-R_{\mathrm{TM}})^2}.
\label{eq:onesample}
\end{equation}
Substituting Equation~\eqref{eq:ko} into Equation~\eqref{eq:onesample} gives the general sample-size equation, for an approximately normally distributed continuous outcome, a single baseline eligibility measurement, and a two-sided test:
\begin{equation}
N = \left\lceil
\frac{\sigma_D^2 (z_{1-\alpha/2}+z_{1-\beta})^2}
{\left[M-(\sigma_1-\rho\sigma_2)\lambda(z)\right]^2}
\right\rceil,
\qquad \text{subject to } M \neq R_{\mathrm{TM}},
\label{eq:softwaregeneral}
\end{equation}
with $\sigma_D$ as in Equation~\eqref{eq:vard}. Because Equation~\eqref{eq:softwaregeneral} depends on $M-R_{\mathrm{TM}}$ only through its square, $M$ may be smaller than $R_{\mathrm{TM}}$; the only value excluded is $M=R_{\mathrm{TM}}$, at which $\Delta_{\mathrm{net}}=0$ and $N$ is undefined (Section~\ref{sec:effectsize}). Under the equal-variance simplification $\sigma_1=\sigma_2=\sigma$, Equation~\eqref{eq:softwaregeneral} reduces to
\begin{equation}
N = \left\lceil
\frac{\sigma^2\left\{(1-\rho^2) + (1-\rho)^2\left[1+z\lambda(z)-\lambda(z)^2\right]\right\}(z_{1-\alpha/2}+z_{1-\beta})^2}
{\left[M-\sigma(1-\rho)\dfrac{\phi(|\kappa-\mu|/\sigma)}{1-\Phi(|\kappa-\mu|/\sigma)}\right]^2}
\right\rceil,
\qquad \text{subject to } M \neq R_{\mathrm{TM}}.
\label{eq:software}
\end{equation}
For a one-sided hypothesis, $z_{1-\alpha/2}$ is replaced by $z_{1-\alpha}$ throughout this section. Equation~\eqref{eq:onesample} can be inverted algebraically to obtain the standard closed-form approximations for power or the minimum detectable effect, as well as for sample size, without numerical search.

\subsection{Power for a given sample size}
\label{sec:power}

For a one-sided test, solving Equation~\eqref{eq:onesample} for $z_{1-\beta}$ and applying $\Phi$ gives the normal-theory power achieved by a given analyzable sample size $n$:
\begin{equation}
1-\beta = \Phi\!\left(\eta - z_{1-\alpha}\right), \qquad \eta = \frac{\sqrt{n}\,\Delta_{\mathrm{net}}}{\sigma_D}.
\label{eq:power-onesided}
\end{equation}
For a two-sided test, Equation~\eqref{eq:power-onesided} with $z_{1-\alpha}$ replaced by $z_{1-\alpha/2}$ is the standard closed-form approximation, obtained by dropping the (usually negligible) probability of rejecting in the direction opposite to $\Delta_{\mathrm{net}}$. That approximation is adequate when $\eta$ is large and positive, the conventional design regime, but it is not symmetric in the sign of $\eta$ and degrades sharply as $\eta$ approaches zero or becomes negative — exactly the regime this method does not exclude, since $\Delta_{\mathrm{net}}<0$ is a valid, signed treatment component (Section~\ref{sec:effectsize}). The two-sided normal-theory power is
\begin{equation}
1-\beta = \left[1-\Phi(z_{1-\alpha/2}-\eta)\right] + \Phi(-z_{1-\alpha/2}-\eta),
\label{eq:power}
\end{equation}
which is symmetric under $\eta\to-\eta$ and reduces to the standard approximation when the second term is negligible. Equation~\eqref{eq:power} evaluates both tails of the assumed normal sampling distribution of $\bar D$ exactly; it is "exact" relative to that normal-theory assumption, not exact finite-sample power under the true, non-normal distribution of $D\mid S$ discussed in Section~\ref{sec:simulation} — a distinction we return to there. The implementation uses Equation~\eqref{eq:power} whenever power is computed directly or reported as the power achieved at a solved sample size; solving for $N$ or the minimum detectable effect (Section~\ref{sec:mde}) retains the standard approximate inversion, since Equation~\eqref{eq:power} has no closed-form inverse in $n$ or $\Delta_{\mathrm{net}}$ — the same practice used elsewhere (for example, closed-form sample-size software commonly solves $N$ approximately while reporting the full normal-theory achieved power). When solving for sample size, the implementation reports the power actually achieved at the rounded integer $N=\lceil n^*\rceil$ from Equation~\eqref{eq:softwaregeneral}, via Equation~\eqref{eq:power} evaluated at $n=N$, rather than the (generally slightly lower) power implied by the unrounded $n^*$.

\subsection{Minimum detectable effect}
\label{sec:mde}

Solving Equation~\eqref{eq:onesample} for $\Delta_{\mathrm{net}}$ given $n$ and a target power $1-\beta$ gives, up to sign, the minimum treatment component detectable with that sample size:
\begin{equation}
|\Delta_{\mathrm{net}}|_{\min} = \frac{\sigma_D\,(z_{1-\alpha/2}+z_{1-\beta})}{\sqrt{n}}.
\label{eq:mde}
\end{equation}
Because Equation~\eqref{eq:onesample} depends on $\Delta_{\mathrm{net}}$ only through its square, both $\Delta_{\mathrm{net}}=+|\Delta_{\mathrm{net}}|_{\min}$ and $\Delta_{\mathrm{net}}=-|\Delta_{\mathrm{net}}|_{\min}$ solve it: an effect of this magnitude is equally detectable whether it points toward the mean (beyond what RTM alone predicts) or away from it (opposing the RTM-driven direction). The implementation reports the positive root, $\Delta_{\mathrm{net}}=+|\Delta_{\mathrm{net}}|_{\min}$, matching the oriented-scale convention of Section~\ref{sec:generalization}, in which positive $\Delta_{\mathrm{net}}$ corresponds to the anticipated treatment direction; the negative alternative is not separately reported but is available by symmetry, $\Delta_{\mathrm{net}}=-|\Delta_{\mathrm{net}}|_{\min}$.

This is one of two related quantities investigators may want from the same calculation: $|\Delta_{\mathrm{net}}|_{\min}$ is the minimum \emph{treatment} component detectable net of RTM, while
\begin{equation}
M = |\Delta_{\mathrm{net}}|_{\min} + R_{\mathrm{TM}}
\label{eq:mde-total}
\end{equation}
(Equation~\eqref{eq:net}) is the corresponding minimum \emph{total} pre-post change an investigator would need to observe, before any RTM adjustment, for that treatment component to be detectable. Neither is more fundamental than the other; they answer different practical questions (“how large must the treatment component be” versus “how large a total change must I actually see”), and the implementation reports both from a single call regardless of which of \texttt{diff()} or \texttt{delta()}, if either, was used to specify the design in the first place — the same pairing already used for $M$ and $\Delta_{\mathrm{net}}$ throughout Section~\ref{sec:software}.

\subsection{Inflation for attrition}

The quantity produced by Equation~\eqref{eq:softwaregeneral} is the number of participants with evaluable pre-post outcome data. If the anticipated attrition proportion is $a \in [0,1)$, the required enrollment size is
\begin{equation}
N_{\mathrm{enroll}} = \left\lceil \frac{N}{1-a} \right\rceil,
\end{equation}
applied after, and separately from, the RTM-adjusted analyzable sample size.

\section{Software Implementation}
\label{sec:software}

We implemented this method via \texttt{power onemean\_rtm}, a community-contributed package for Stata \citep{linden2026poweronemeanrtm}. The program computes closed-form solutions for sample size, power, and minimum detectable effect using the algebraic inversions of Equation~\eqref{eq:onesample} described above.

Beyond the core parameters ($\mu$, $\sigma_1$, $\sigma_2$, $\rho$, $\kappa$), the implementation distinguishes two ways of specifying the target effect:
\[
\texttt{diff()} \longleftrightarrow M, \qquad \texttt{delta()} \longleftrightarrow \Delta_{\mathrm{net}}.
\]
This distinction matters in practice: when the eligibility cutoff is far into the tail of the baseline distribution, RTM can account for the great majority of an anticipated raw change, leaving a small, easily overlooked $\Delta_{\mathrm{net}}$ and a correspondingly large required sample size that correctly reflects how little detectable signal remains — not a computational error, but the central substantive point of the method. Investigators who already know the effect size they wish to detect net of RTM may specify it directly via \texttt{delta()} rather than back-solving for the equivalent raw $M$. Only one of \texttt{diff()}/\texttt{delta()} may be supplied as input, but both $M$ and $\Delta_{\mathrm{net}}$ are always reported as output, including when neither is supplied and both are instead solved for as the minimum detectable effect (Section~\ref{sec:mde}).

The implementation additionally supports an optional follow-up standard deviation -- denoted \texttt{sd2} here, paired with \texttt{sd1} for the baseline SD $\sigma_1$ (see the Appendix for the corresponding Stata option names) -- entering both the RTM term via Equation~\eqref{eq:ko} and the change-score variance via Equation~\eqref{eq:vard}; when \texttt{sd2} is not specified, both reduce automatically to their equal-variance forms, Equations~\eqref{eq:rtm} and~\eqref{eq:sdchange}. The implementation uses the cutoff-selection-conditional variance throughout (Section~\ref{sec:vard}), matching the derivation above rather than the unconditional alternative discussed and ruled out there. Because Equation~\eqref{eq:ko} permits a negative RTM term when $\rho\sigma_2>\sigma_1$ (Section~\ref{sec:generalization}), and because $\Delta_{\mathrm{net}}<0$ is itself a valid signed treatment component (Section~\ref{sec:effectsize}), the implementation prints a non-fatal note in either case rather than treating it as an error, since the downstream algebra remains valid; Section~\ref{sec:example} gives worked examples of both. Only $M=R_{\mathrm{TM}}$ exactly is rejected, since $\Delta_{\mathrm{net}}=0$ leaves no effect to power. The implementation also supports an attrition proportion (\texttt{dropout()}, producing a reported enrollment size distinct from the analyzable sample size) and a one-sided/two-sided test option; achieved power is computed using the full two-sided normal-theory expression in Equation~\eqref{eq:power}, rather than the one-tail approximation, as described in Section~\ref{sec:power}. Because every method-specific numeric input accepts a Stata \texttt{numlist}, sensitivity analysis across any combination of parameters is available through \texttt{power}'s native table and graph machinery without additional custom code; a combination for which $M=R_{\mathrm{TM}}$ exactly is reported as missing within a multi-value sweep rather than aborting the computation for other, valid combinations.

\section{Illustrative Example}
\label{sec:example}

Stata code reproducing every example in this section is given in the Appendix.

Consider a design with baseline population mean $\mu=50$, baseline SD $\sigma=10$, and pre-post correlation $\rho=0.6$. A moderate eligibility cutoff of $\kappa=60$ (one baseline SD above the mean) and an anticipated total change of $M=15$ gives, via Equation~\eqref{eq:rtm},
\[
z = \frac{|60-50|}{10} = 1.0, \qquad
\lambda(1.0) \approx 1.525, \qquad
R_{\mathrm{TM}} = 10(1-0.6)(1.525) \approx 6.101,
\]
so RTM accounts for about 41\% of the anticipated total change, leaving $\Delta_{\mathrm{net}} = 15 - 6.101 \approx 8.899$. Using Equation~\eqref{eq:sdchange}, $\sigma_D\approx8.197$, and Equation~\eqref{eq:software} gives $N=7$ for 80\% power at a two-sided $\alpha=0.05$. A conventional calculation that ignored RTM entirely — using the same $\sigma_D$ but powering the full $M=15$ against a null of zero, rather than $\Delta_{\mathrm{net}}$ against the RTM-implied null — would instead report $N=3$: less than half the RTM-adjusted requirement, and underpowered for the effect actually of scientific interest.

We next consider a deliberately extreme case, to illustrate the boundary behavior as the anticipated change approaches what RTM alone predicts. Retaining $\mu=50$, $\sigma=10$, $\rho=0.6$ but returning to $\kappa=65$ (1.5 baseline SDs above the mean) and $M=8$:
\[
z = 1.5, \qquad \lambda(1.5)\approx1.939, \qquad R_{\mathrm{TM}} = 10(1-0.6)(1.939) \approx 7.755,
\]
so that $\Delta_{\mathrm{net}} = 8 - 7.755 = 0.245$: here RTM accounts for roughly 97\% of the anticipated change, and only a small residual remains to detect. Using Equation~\eqref{eq:sdchange}, $\sigma_D\approx8.148$, and Equation~\eqref{eq:software} gives $N=8{,}661$ for the same power and significance level — versus a conventional $N=9$ if the full $M=8$ were powered against a null of zero, ignoring RTM. The enormous gap between these two numbers is the point of the example, not an artifact of it: as $M\to R_{\mathrm{TM}}$, $\Delta_{\mathrm{net}}\to0$ and $N\to\infty$ (Section~\ref{sec:effectsize}), so a design in which nearly all of the anticipated change is attributable to RTM genuinely does require an enormous sample to detect the small remaining signal reliably, and reporting a modest conventional $N$ for such a design would be materially misleading.

By contrast, specifying the same extreme scenario with the RTM-adjusted effect given directly, $\Delta_{\mathrm{net}}=8$ (i.e., \texttt{delta(8)} rather than \texttt{diff(8)}), corresponds to a raw anticipated change of $M = 8 + 7.755 = 15.755$ and requires $N=9$ for the same power and significance level — illustrating both the magnitude of the RTM adjustment in this cutoff-selected design and the practical importance of the \texttt{diff}/\texttt{delta} distinction described in Section~\ref{sec:software}.

\subsection{Unequal variance and a sign-reversing example}

Retaining $\mu=50$, $\sigma_1=10$, $\rho=0.6$, $\kappa=65$, and $M=8$, now suppose follow-up variance exceeds baseline variance, $\sigma_2=15$. Applying Equation~\eqref{eq:ko}:
\[
R_{\mathrm{TM}} = (10 - 0.6\times15)(1.939) \approx 1.939, \qquad
\Delta_{\mathrm{net}} = 8 - 1.939 = 6.061,
\]
a materially smaller RTM adjustment than the equal-variance case above. Using Equation~\eqref{eq:vard}, $\sigma_D \approx 12.006$, and Equation~\eqref{eq:softwaregeneral} gives $N=31$ for the same power and significance level as before — far smaller than the equal-variance case's $N=8{,}661$, illustrating that the direction of the variance inequality here ($\sigma_2>\sigma_1$) both attenuates the RTM term and changes the change-score variance substantially.

Pushing further, with $\rho=0.9$ and $\sigma_2=20$:
\[
R_{\mathrm{TM}} = (10 - 0.9\times20)(1.939) \approx -15.509, \qquad
\Delta_{\mathrm{net}} = 8 - (-15.509) = 23.509.
\]
RTM is negative: given this baseline variance, follow-up variance, and correlation, the model implies that participants selected on an extreme baseline value are expected to move \emph{further from} the population mean on remeasurement, not toward it. The resulting $\Delta_{\mathrm{net}}$ exceeds $M$ itself, since a negative quantity is being subtracted; here $\sigma_D\approx9.250$ and the closed-form calculation returns $N=2$. This number should not be read as a literal enrollment recommendation: the large-sample normal approximation underlying Equation~\eqref{eq:onesample} is not credible at $N=2$, and a result this small indicates a very large standardized effect relative to $\sigma_D$ rather than a trustworthy sample-size target. We report it here because it is what the closed-form calculation returns, and because a negative RTM this large is itself a signal worth scrutinizing (Section~\ref{sec:discussion}) rather than accepting mechanically. Consistent with the recommendation in Section~\ref{sec:discussion}, a design landing this far below the approximate $N<16$ guideline should have its achieved power confirmed by direct simulation rather than taken from Equation~\eqref{eq:onesample} alone. Both examples are numerically confirmed against an independent implementation and reduce exactly to the corresponding equal-variance results when $\sigma_2=\sigma_1$.

\subsection{A negative treatment component, and why two-sided normal-theory power matters}
\label{sec:negdelta}

Returning to the equal-variance case ($\mu=50$, $\sigma=10$, $\rho=0.6$, $\kappa=65$, $R_{\mathrm{TM}}\approx7.755$), now suppose the investigator anticipates a smaller total change, $M=5$, which is \emph{less} than $R_{\mathrm{TM}}$:
\[
\Delta_{\mathrm{net}} = 5 - 7.755 = -2.755.
\]
This is a negative treatment component: the anticipated total change falls short of what RTM alone predicts, meaning the design anticipates an effect \emph{opposing} the RTM-driven direction. At $n=150$, Equation~\eqref{eq:power} gives a two-sided normal-theory power of $0.985$ for detecting this effect. The standard one-tail approximation, evaluated at the same $\eta=\sqrt{n}\,\Delta_{\mathrm{net}}/\sigma_D\approx-4.14$, gives $0.000$ — not a small underestimate, but a complete failure, since the approximation is not symmetric in the sign of $\eta$ and collapses to zero for the more extreme negative values this scenario produces. Solving for the sample size needed to detect this same effect at 80\% power (using the standard approximate inversion, Equation~\eqref{eq:softwaregeneral}) gives $N=69$, which achieves a two-sided normal-theory power of $0.802$ when checked against Equation~\eqref{eq:power} — confirming that the approximate inversion used for sample-size solving remains accurate even though the one-tail power approximation itself is not.

\section{Simulation Study}
\label{sec:simulation}

The illustrative examples above show what the method computes; they do not by themselves establish that the closed-form $N$ delivers the power it claims, or that the unusual RTM-implied null is tested correctly. We address both questions directly by simulation. We refer to this as the \emph{operating-characteristics simulation}, distinct from the variance-validation simulation of Section~\ref{sec:validation}, which checked a single intermediate formula rather than the sample-size method's end-to-end behavior.

\subsection{Design}

For a given scenario ($\mu$, $\sigma_1$, $\sigma_2$, $\rho$, $\kappa$, $\alpha$, and either a target power or a true treatment component), we (1) compute the closed-form $N$ from Equation~\eqref{eq:softwaregeneral} (power scenarios) or take $N$ from a matching power scenario (Type I error scenarios, described below); (2) simulate $B$ independent studies under the exact assumed data-generating process — draw $N$ baseline values from the cutoff-truncated normal distribution via exact inverse-CDF sampling (not rejection sampling), generate paired follow-up values from the regression decomposition of Section~\ref{sec:vard} with the scenario's true treatment component imposed on the oriented scale of Section~\ref{sec:generalization}, and run the two-sided test of $H_0:\mu_D=R_{\mathrm{TM}}$ using the known population $\sigma_D$ (a z-test, matching the paper's own large-sample derivation exactly rather than mixing in sample-variance estimation as a separate source of error); and (3) record the empirical rejection rate, its Monte Carlo standard error $\sqrt{\hat p(1-\hat p)/B}$, and a 95\% Wilson score confidence interval.

Two families of scenarios were run, 58 in total. \emph{Power scenarios} (42) cross cutoff extremity ($z\in\{0.5,1.0,1.5,2.0\}$), correlation ($\rho\in\{0.3,0.6,0.9\}$), and the sign of the true treatment component (in the same direction as the RTM-driven change, or opposing it), at equal variance and 80\% target power; a secondary grid varies the follow-up/baseline variance ratio ($\sigma_2/\sigma_1\in\{0.5,1.5,2.0\}$), including the paper's own negative-RTM combination (Section~\ref{sec:negdelta}); a further set spot-checks target power at 0.5 and 0.9; the paper's own two worked examples (Section~\ref{sec:example}) are included with their exact parameters; and four scenarios check lower-tail selection specifically, to confirm the tail-symmetry derivation of Section~\ref{sec:generalization} holds in simulation and not merely in algebra. \emph{Type I error scenarios} (16) reuse the ($z$, $\rho$) and variance-ratio combinations from the power grid, at the $N$ a matching power scenario implied, but with no true treatment component in the data-generating process — testing whether the empirical rejection rate matches nominal $\alpha=0.05$ under the RTM-only null, independent of whether the power calculation is also accurate. $B$ was scaled inversely with $N$ (50{,}000 down to 5{,}000) to bound total computation while preserving adequate precision; at $B=50{,}000$ the Monte Carlo SE on a proportion near 0.80 is approximately 0.0018 — precise enough that even a difference of a percentage point or two between target and achieved power is statistically detectable, a point we return to below.

\subsection{Results}

\textbf{Type I error.} Empirical Type I error was tightly centered on the nominal 0.05 level across all 16 scenarios (range $0.0489$--$0.0522$; mean signed difference $0.0002$), with no discernible pattern by cutoff, correlation, selection tail, variance ratio, or sample size. Fifteen of 16 Wilson intervals contained 0.05; with 16 95\% intervals, occasional noncoverage is expected even under correct calibration, so this is better read as confirmation than as a near-failure. Full results are in Supplement Table 2 (see the accompanying Supplement).

\textbf{Power.} Empirical power showed a clear sample-size-dependent pattern (Supplement Table 1). At very small sample sizes ($N\le15$), the closed-form calculation was conservative, with achieved power exceeding target power by an average of $0.041$. The discrepancy declined rapidly as sample size increased: mean differences were $0.0067$ for $16\le N\le60$, $0.0021$ for $61\le N\le200$, and $-0.0050$ at the paper's own extreme example ($N=8{,}661$). Restricting to the 16 scenarios with $N\ge16$, mean absolute error was $0.0062$ — under one percentage point — confirming that the small-$N$ pattern, not a general miscalibration, accounts for essentially all of the discrepancy seen when every scenario is pooled together. Because $B=50{,}000$ gives Monte Carlo precision fine enough to detect even a one-point difference as statistically significant, we do not lean on the fraction of scenarios whose CI contains the target (7 of 42) as the headline result; that fraction would look far worse than the substantive conclusion warrants purely because the intervals are so narrow. The pattern by sample size, and the resulting absolute error, are the more informative summary.

This small-$N$ conservatism has a specific explanation grounded in the same derivation as the rest of the method, not merely a generic caveat about normal approximations. The two-sided power formula, Equation~\eqref{eq:power}, is exact given that $\bar D$, the sample mean of the change score among cutoff-selected participants, is normally distributed — but $D\mid S$ itself is not normal, even though $(Y_0,Y_1)$ is bivariate normal: $Y_0\mid S$ is a truncated normal, so $D\mid S=(\beta-1)(Y_0-\mu)+\varepsilon$ (Section~\ref{sec:vard}) inherits that truncation. Normality of $\bar D$ is therefore asymptotic, guaranteed by the central limit theorem as $N$ grows, rather than exact at any fixed $N$. This is a direct, expected consequence of cutoff selection — the same feature the method is built to handle — rather than a defect specific to the RTM adjustment; we do not, however, have a non-RTM comparator simulation to formally isolate finite-sample bias attributable to this truncation from finite-sample bias any ordinary normal-approximation power formula would show at comparably small $N$, and we do not claim more than the pattern in Supplement Table 1 directly supports.

This is direct empirical support for the caution already given qualitatively for the $N=2$ and $N=3$ results in Section~\ref{sec:negdelta} and Supplement Table 3: such results should be read as indicating a very large standardized effect rather than as precise power guarantees, while results at $N\ge16$ or so can be read considerably more literally, per the aggregate error above. One caveat on this grid's design: because $\rho$ and the delta-magnitudes used mechanically produce smaller $N$ together, the apparent association between high $\rho$ and larger bias in Supplement Table 3 is substantially explained by $N$ rather than necessarily an independent effect of $\rho$; a grid holding $N$ fixed while varying $\rho$ was not run and would be needed to fully separate the two. Full scenario-by-scenario results are in Supplement Table 3 (see the accompanying Supplement).

\section{Discussion}
\label{sec:discussion}

The principal result is that RTM affects prospective sample-size determination through two distinct mechanisms. First, expected RTM reduces the treatment component of an anticipated total pre-post change from $M$ to $\Delta_{\mathrm{net}}=M-R_{\mathrm{TM}}$ (Section~\ref{sec:effectsize}). Second, cutoff-based selection changes the variance of the pre-post difference, requiring $\mathrm{Var}(D\mid S)$ rather than the unconditional change-score variance (Section~\ref{sec:vard}). Both quantities are determined by the same underlying baseline/follow-up distribution and eligibility rule; indeed, the same combination $(\sigma_1-\rho\sigma_2)$ enters both the RTM term and the conditional variance (Section~\ref{sec:vard}), so the two adjustments are not independent corrections bolted onto a conventional calculation, but two consequences of the same cutoff-selected sampling process.

Simulation provided two complementary checks on this framework (Section~\ref{sec:simulation}). First, empirical Type I error remained closely centered on the nominal 0.05 level under the RTM-only null, supporting the use of the RTM-implied counterfactual as the hypothesis actually being tested. Second, the closed-form sample-size calculation tended to over-deliver rather than under-deliver target power, with the discrepancy concentrated at very small sample sizes and diminishing rapidly as $N$ increased, to well under a percentage point by $N\ge16$. This behavior is consistent with the non-normal finite-sample distribution of the change score induced by cutoff selection: although the source measurements are bivariate normal, conditioning on an extreme baseline value truncates the baseline distribution, so normality of the sample mean $\bar D$ is asymptotic rather than exact at any fixed $N$ — a direct consequence of the same selection mechanism the method exists to handle, not an incidental numerical artifact. A practical recommendation follows directly: in our simulations, discrepancies were concentrated at $N\le15$ (Supplement Table 1). We therefore recommend confirming achieved power by direct Monte Carlo simulation, following the approach of this section, when the closed-form calculation produces a very small sample size (approximately $N<16$), particularly when the cutoff is extreme or the standardized treatment component is large; at larger $N$ the closed-form figure can be trusted with little adjustment. This threshold reflects how our simulation results happened to be grouped rather than a theoretical discontinuity at $N=16$ specifically.

A conventional one-mean power calculation for this design would use $\Delta_{\mathrm{conventional}}=M$ rather than $\Delta_{\mathrm{net}}=M-R_{\mathrm{TM}}$. Because $\Delta$ enters Equation~\eqref{eq:onesample} squared, even a moderate RTM adjustment can produce a large change in the required $N$: the moderate example in Section~\ref{sec:example} gives a conventional $N$ less than half the RTM-adjusted requirement, and the extreme example gives a gap of several orders of magnitude. This is probably the single most practically important implication of the method: a conventional pre-post sample-size calculation for a cutoff-selected population is not merely imprecise, but can be misleading in a specific, predictable direction, systematically understating the sample needed to detect the treatment component net of expected RTM.

The method complements rather than replaces the existing RTM literature reviewed in the Introduction. \citet{khanolivier2025} provide the decomposition and general distributional framework underlying $\Delta_{\mathrm{net}}=M-R_{\mathrm{TM}}$; the present method uses that framework prospectively rather than retrospectively. Likewise, \citet{mcmahon1994} and \citet{goldenholz2023} establish that RTM is relevant to sample-size planning in other outcome structures and through simulation; the contribution here is a closed-form solution specific to the continuous, cutoff-selected, single-arm setting. None of this displaces the more basic point, already noted in the Introduction, that RTM adjustment addresses one specific threat to validity in an uncontrolled design and not others \citep{campbellstanley1966,shadishcookcampbell2002}; the method is offered for settings in which a single-arm design is already being contemplated for other reasons, not as a substitute for randomization or a concurrent control, and in that circumstance ignoring a predictable source of apparent change at the design stage is worse than accounting for it.

Several further limitations warrant explicit discussion. First, like the retrospective methods it builds on, this approach assumes an approximately normal baseline distribution and an approximately bivariate-normal baseline/follow-up relationship. Linden's (\citeyear{linden2013}) simulations found that the conventional RTM equations, accurate under multivariate normality, substantially underestimated RTM in an empirical example involving skewed data; the same caution applies here, and the method should be regarded as a normal-theory tool pending simulation-based or transformation-based extensions for non-normal outcomes.

Second, the present derivation treats $\mu$, $\sigma_1$, $\sigma_2$, $\rho$, and $\kappa$ as fixed design parameters, whereas in practice they are typically estimated from historical, pilot, or published data and are themselves subject to sampling uncertainty. Formal propagation of that uncertainty into the sample-size calculation (for example, via confidence intervals on the RTM estimate, as considered separately for the retrospective estimation setting) is not addressed here; investigators can, however, explore its practical consequences using the implementation's sensitivity analysis (Section~\ref{sec:software}), by supplying a plausible range of values for any input parameter and observing how $N$, power, or the minimum detectable effect respond, rather than relying on a single point estimate of each parameter.

Third, the method assumes an investigator can meaningfully specify $M$ — or, via \texttt{delta()}, $\Delta_{\mathrm{net}}$ directly — and this is not always straightforward. Historical estimates of pre-post change from similarly cutoff-selected populations may already contain RTM and so correspond more naturally to $M$, whereas treatment-effect estimates from controlled studies may more closely represent $\Delta_{\mathrm{net}}$ directly. Investigators should match the command input to the provenance of the effect-size estimate they have in hand, rather than defaulting to \texttt{diff()} out of habit; this is precisely why the implementation supports both.

Fourth, because $\sigma_1$, $\sigma_2$, and $\rho$ may be specified independently, some combinations imply unusual behavior, including negative RTM (Section~\ref{sec:generalization}) or the very small closed-form sample sizes illustrated in Section~\ref{sec:example}. Such combinations are permitted by the general model and the downstream algebra remains valid, but their occurrence should prompt investigators to examine whether the assumed follow-up variance and pre-post correlation are actually plausible for the intended population, rather than being accepted merely because the formula returns a numerical answer.

Fifth, the current implementation assumes a single baseline eligibility measurement; Linden's (\citeyear{linden2013}) extension to eligibility based on the mean of multiple baseline measurements, which reduces within-subject variability and therefore reduces RTM, is a natural but distinct extension left for future work.

Sixth, while Equation~\eqref{eq:ko} relaxes the equal-variance assumption within the RTM term, both it and our sample-size derivation retain \citet{khanolivier2025}'s restriction to the bivariate normal case; their more general result, applicable to any bivariate distribution, is not incorporated here and would require distribution-specific re-derivation of Equation~\eqref{eq:onesample} rather than a direct substitution.

\section{Conclusion}

Selecting participants on an extreme baseline value creates an expected pre-post change even in the absence of treatment. We derived a closed-form prospective sample-size method that accounts for this in two ways: by powering the treatment component remaining after expected RTM is removed from the anticipated total change, and by using the variance of the pre-post change conditional on the same cutoff-based eligibility rule. The method extends established cutoff-based RTM methodology to prospective planning for continuous single-arm pre-post studies and accommodates unequal baseline and follow-up variances. Monte Carlo simulation demonstrated accurate Type I error control and close agreement between nominal and achieved power at conventional sample sizes, while identifying very small samples as a setting in which direct simulation is advisable. Its implementation in \texttt{power onemean\_rtm} provides closed-form solutions for sample size, power, and minimum detectable effect, together with attrition adjustment and sensitivity analysis. The resulting calculations make explicit when an apparently meaningful anticipated pre-post change contains little treatment signal once the change expected from RTM is taken into account.

\bibliographystyle{apalike}
\bibliography{rtm_onemean_paper}

\clearpage
\section*{Appendix: Stata Code for Illustrative Examples}
\label{sec:appendix-code}

\addcontentsline{toc}{section}{Appendix: Stata Code for Illustrative Examples}

The commands below reproduce all worked examples in Section~\ref{sec:example}, and can be copied directly into Stata (with \texttt{power onemean\_rtm} installed) to reproduce the reported results. Note on syntax: the paper refers to the baseline and follow-up standard deviations as \texttt{sd1} and \texttt{sd2} (matching $\sigma_1$ and $\sigma_2$). The corresponding Stata options are \texttt{sd()} and \texttt{sd1()}, respectively.

\subsection*{Moderate example (Section~\ref{sec:example})}
\begin{verbatim}
. power onemean_rtm, mu(50) sd(10) corr(0.6) cutoff(60) diff(15) power(0.8)
\end{verbatim}

\subsection*{Extreme example, specified via \texttt{diff()} (Section~\ref{sec:example})}
\begin{verbatim}
. power onemean_rtm, mu(50) sd(10) corr(0.6) cutoff(65) diff(8) power(0.8)
\end{verbatim}

\subsection*{Same extreme example, specified via \texttt{delta()} instead (Section~\ref{sec:example})}
\begin{verbatim}
. power onemean_rtm, mu(50) sd(10) corr(0.6) cutoff(65) delta(8) power(0.8)
\end{verbatim}

\subsection*{Unequal variance, $\sigma_2=15$ (Section~\ref{sec:example}, ``Unequal variance and a sign-reversing example'')}
\begin{verbatim}
. power onemean_rtm, mu(50) sd(10) corr(0.6) cutoff(65) diff(8) sd1(15) power(0.8)
\end{verbatim}

\subsection*{Negative RTM, $\rho=0.9$, $\sigma_2=20$ (Section~\ref{sec:example}, same subsection)}
\begin{verbatim}
. power onemean_rtm, mu(50) sd(10) corr(0.9) cutoff(65) diff(8) sd1(20) power(0.8)
\end{verbatim}

\subsection*{Negative treatment component (Section~\ref{sec:negdelta})}
Achieved power at $n=150$:
\begin{verbatim}
. power onemean_rtm, mu(50) sd(10) corr(0.6) cutoff(65) diff(5) n(150)
\end{verbatim}
Sample size needed for 80\% power at this same effect:
\begin{verbatim}
. power onemean_rtm, mu(50) sd(10) corr(0.6) cutoff(65) diff(5) power(0.8)
\end{verbatim}

\end{document}